\documentclass[twocolumn,amsmath,amssymb,preprintnumbers,pra]{revtex4-1}

\usepackage[dvipdfmx]{graphicx}
\usepackage{dcolumn}
\usepackage{bm}
\usepackage{color}

\begin{document}

\title{Excited-State Adiabatic Quantum Computation Started with Vacuum States}

\author{Hayato Goto and Taro Kanao}
\affiliation{
Frontier Research Laboratory, 
Corporate Research \& Development Center, 
Toshiba Corporation, 
1, Komukai Toshiba-cho, Saiwai-ku, Kawasaki-shi, 212-8582, Japan}

\date{\today}

\begin{abstract}
Adiabatic quantum computation (AQC),
which is particularly useful for combinatorial optimization, 
becomes more powerful by using excited states, instead of ground states.
However, the excited-state AQC is prone to errors due to dissipation.
Here we propose the excited-state AQC started with
the most stable state, i.e., the vacuum state.
This counterintuitive approach becomes possible by using a driven quantum system,
or more precisely, a network of Kerr-nonlinear parametric oscillators (KPOs).
By numerical simulations, we show that some hard instances, 
where standard ground-state AQC with KPOs fails to find their optimal solutions, 
can be solved by the present approach,
where nonadiabatic transitions are rather utilized.
We also show that the use of the vacuum state as an initial state leads to robustness 
against errors due to dissipation,
as expected, compared to the use of a really excited (nonvacuum) state as an initial state.
Thus, the present work offers new possibilities for quantum computation and driven quantum systems.

\end{abstract}

\maketitle

\section{Introduction}

Adiabatic quantum computation (AQC)~\cite{Farhi2000a,Farhi2001a,Albash2018a} or 
quantum annealing~\cite{Kadowaki1998a,Santoro2002a,Das2008a}
is an alternative approach to quantum computation.
The AQC is particularly useful for combinatorial optimization problems,
where we have to minimize (or maximize) functions of discrete variables called objective (or cost) functions~\cite{Metaheuristics}.
The Ising problem (search for ground states of Ising spin models)~\cite{Barahona1982a,Lucas2014a}
is a typical example of such problems.
Ising machines, which are designed for solving the Ising problem, 
based on AQC are expected to be useful for practical applications,
because there are various situations requiring to solve combinatorial optimization 
problems which can be mapped 
to the Ising problem~\cite{Barahona1988a,Sakaguchi2016a,Rosenberg2016a}.

The idea of AQC is simple.
We start with the ground state of an initial Hamiltonian,
where we know the ground state because the initial Hamiltonian is simple enough.
Changing the Hamiltonian slowly to the one 
corresponding to the objective function for a given problem,
we finally obtain the ground state of the final Hamiltonian 
assuming that the quantum adiabatic theorem~\cite{Messiah} holds.
The final ground state gives us the solution of the problem.
There is, however, a problem. 
If the energy gap between the ground and first excited states almost closes
during the AQC,
the adiabatic theorem does not hold,
and consequently we cannot find the ground state. 
Thus, the energy-gap closing is a fatal problem for AQC.

One of the approaches to this crucial problem
is to use excited states in AQC.
For instance,
we can achieve an exponential quantum speedup 
by using an excited state via nonadiabatic transitions 
at energy-gap closing points during AQC~\cite{Somma2012a}.
Moreover, it is known that 
``stoquastic" AQC~\cite{Albash2018a,Bravyi2008a}, 
to which we can apply a classical simulation method~\cite{Santoro2002a},
becomes as powerful as universal quantum computation
by using excited states~\cite{Albash2018a,Jordan2010a}.
That is, the use of excited states makes AQC more powerful.
The positive use of excited states in AQC, 
where the initial state is intentionally set to an excited state, not the ground state, 
has been proposed~\cite{Crosson2014a}.
However, this approach is accompanied by the problem 
that the initial state is prone to errors due to dissipation.

In this paper, 
we propose a new approach to the excited-state AQC,
where the initial state is the most stable state, i.e., the vacuum state.
This counterintuitive approach is possible by using a driven quantum system,
or more precisely, a network of Kerr-nonlinear parametric oscillators (KPOs).
The concept of the KPO and 
quantum computations with KPOs, 
both the AQC and gate-based universal quantum computation, 
were proposed in Refs.~\citenum{Goto2016a}
and \citenum{Goto2016b},
which were discovered inspired 
by a more classical approach using 
optical parametric oscillators~\cite{Goto2019b,Wang2013a,Marandi2014a,Yamamoto2017a}.
The proposals have been followed by 
interesting related works, such as 
superconducting-circuit implementations of KPOs, 
generation of Schr\"odinger cat states using KPOs,
and theoretical studies on KPOs as new driven quantum  systems~\cite{Bartolo2016a,Zhang2017a,Puri2017a,Nigg2017a,
Puri2017b,Savona2017a,Zhao2018a,Goto2018a,Dykman2018a,Goto2019a,Rota2019a,
Wang2019a,Puri2019a,Grimm2019a,Puri2019b,Teh2020a,
Verstraelen2020a,Roberts2020a,Kewming2020a}.
The characteristic feature of the AQC with KPOs
is the use of the \textit{effective} Hamiltonian for the driven system,
which enables the vacuum state to be an excited state of the Hamiltonian used for AQC.
This intriguing property of driven systems was exploited for preparing quasienergy excited states
of a KPO via quantum adiabatic evolution started with the vacuum state~\cite{Zhang2017a}.
This is the essential point for our approach.

This paper is organized as follows.
In Sec.~\ref{sec-ground-AQC},
we briefly describe standard ground-state AQC with KPOs,
and show simulation results in order to clarify the energy-gap closing problem.
In Sec.~\ref{sec-excited-AQC},
we explain the proposed approach, that is, 
the excited-state AQC started with vacuum states,
and demonstrate the usefulness of this approach.
In Sec.~\ref{sec-dissipation},
we examine the effects of dissipation using numerical simulations,
where AQC started with a really excited (nonvacuum) state is also simulated for comparison.
In Sec.~\ref{sec-conclusion}, we briefly summarize the present work.

\section{Ground-state AQC with KPOs}
\label{sec-ground-AQC}

The $N$-spin Ising problem with coupling coefficients $\{ J_{i,j} \}$
and local fields $\{ h_i \}$ is to find a spin configuration minimizing the (dimensionless) Ising energy
defined as
\begin{align}
E_{\mathrm{Ising}}
=-\frac{1}{2} \sum_{i=1}^N \sum_{j=1}^N J_{i,j} s_i s_j - \sum_{i=1}^N h_i s_i,
\label{eq-Ising}
\end{align}
where $s_i$ is the $i$th spin taking 1 or $-1$, and the coupling coefficients satisfy
${J_{i,j}=J_{j,i}}$ and ${J_{i,i}=0}$.

The standard ground-state AQC with KPOs is as follows.
To solve the Ising problem, 
we use a KPO network defined by the following Hamiltonian~\cite{Goto2016a,Goto2018a,Goto2019b,comment-detuning}:
\begin{align}
H(t)
&=
\hbar \sum_{i=1}^N 
\left[ \frac{K}{2} a_i^{\dagger 2} a_i^2 +\Delta_i \! (t) a_i^{\dagger} a_i
- \frac{p(t)}{2} \! \left( a_i^2 + a_i^{\dagger 2} \right) \right]
\nonumber
\\
&+\hbar \xi (t) 
\! \left[
-\sum_{i=1}^N \sum_{j=1}^N J_{i,j} a_i^{\dagger} a_j 
- A(t) \sum_{i=1}^N h_i \! \left( a_i + a_i^{\dagger} \right)
\right] \!,
\label{eq-H}
\end{align}
where $\hbar$ is the reduced Planck constant,
$a_i$ and $a_i^{\dagger}$ are the annihilation and creation operators, respectively,
for the $i$th KPO, 
$K$ is the Kerr coefficient,
$\Delta_i \! (t)$ is the detuning frequency for the $i$th KPO,
$p(t)$ is the parametric pump amplitude, 
and $\xi (t)$ and $A(t)$ are control parameters~\cite{comment-xi-A}.
In this work, we assume that all the parameters are positive 
 (if not mentioned)~\cite{comment-parameters}.
Note that the above Hamiltonian is an \textit{effective} one 
in a frame rotating at half the pump frequency and in the rotating-wave approximation~\cite{Goto2016a,Goto2019b}.

We increase $p(t)$ from zero to a sufficiently large value $p_f$ (larger than $K$), 
decrease $\Delta_i \! (t)$ from $\Delta_i^{(0)}$ to zero, 
increase $\xi (t)$ from zero to a small value $\xi_f$ (smaller than $K$),
and set $A(t)$ as ${A(t)=\sqrt{p(t)/K}}$.
Then, the initial and final Hamiltonians, $H_0$ and $H_f$, become
\begin{align}
H_0
&=
\hbar \sum_{i=1}^N 
\! \left( \frac{K}{2} a_i^{\dagger 2} a_i^2 +\Delta_i^{(0)} a_i^{\dagger} a_i \right) \!,
\label{eq-H0}
\\
H_f
&=
\hbar \frac{K}{2} \sum_{i=1}^N 
\! \left( a_i^{\dagger 2} - \alpha_f^2 \right)
\left( a_i^2 -\alpha_f^2 \right)
\nonumber
\\
&+\hbar \xi_f 
\! \left[
-\sum_{i=1}^N \sum_{j=1}^N J_{i,j} a_i^{\dagger} a_j 
- \alpha_f \sum_{i=1}^N h_i \! \left( a_i + a_i^{\dagger} \right)
\right] \!,
\label{eq-Hf}
\end{align}
where ${\alpha_f = \sqrt{p_f/K}}$ and 
a constant term, ${-\hbar K \alpha_f^4/2}$, has been dropped in Eq.~(\ref{eq-Hf}).

From Eq.~(\ref{eq-H0}), 
we find that the initial ground state is exactly the vacuum state.
On the other hand, the first term of $H_f$, 
which is positive semidefinite, 
has the degenerate ground states expressed as
tensor products of coherent states with amplitudes ${\pm \alpha_f}$,
$|{\pm \alpha_f} \rangle \cdots |{\pm \alpha_f} \rangle$~\cite{comment-coherent}.
Thus, assuming a sufficiently small $\xi_f$ compared to $K$,
the final ground state  is approximately given by 
$|{s_1 \alpha_f} \rangle \cdots |{s_N \alpha_f} \rangle$,
where $\{ {s_i=\pm 1} \}$ minimizes the following energy:
\begin{align}
E_f
=2 \hbar \xi_f \alpha_f^2
\! \left(
- \frac{1}{2} \sum_{i=1}^N \sum_{j=1}^N J_{i,j} s_i s_j 
- \sum_{i=1}^N h_i s_i \right) \!.
\label{eq-Ef}
\end{align}
Importantly, this is proportional to the Ising energy in Eq.~(\ref{eq-Ising}):
${E_f \propto E_{\mathrm{Ising}}}$.
Consequently, 
we can obtain the solution of the Ising problem 
from the final state of the adiabatic evolution started with the vacuum state,
assuming that $p(t)$ varies sufficiently slowly and
the quantum adiabatic theorem holds.

To evaluate the ground-state AQC with KPOs,
we solved 1000 random instances of the four-spin Ising problem,
where we numerically solved the Schr\"odinger equation with the Hamiltonian in Eq.~(\ref{eq-H})
(see Appendix~\ref{sec-simulation} for the details of the simulation). 
The results are shown by the histograms in Figs.~\ref{fig-histogram}(a) and \ref{fig-histogram}(b), 
where the failure probability is the probability that we fail to obtain 
the ground state of the Ising problem and 
the residual energy is the difference between the ground-state energy of the Ising problem 
and the expectation value of the Ising energy obtained by the AQC~\cite{comment-probability}.
It is found that most instances are well solved by the ground-state AQC.

To magnify bad results, we plot the results in a two-dimensional plane, 
as shown in Fig.~\ref{fig-histogram}(c).
It turns out that the ground-state AQC results in high failure probabilities or 
high residual energies in some instances.
To examine the reason,
we check the energy levels in a bad instance 
indicated by the vertical arrow in Fig.~\ref{fig-histogram}(c).
(The details of this instance are provided in Appendix~\ref{sec-instance} and 
the numerical calculation of the energies is explained in Appendix~\ref{sec-energy}.)
As shown in Fig.~\ref{fig-histogram}(d),
the energy gap between the ground and first excited states almost closes
at the point indicated by the vertical arrow.
Thus, the reason for the bad result in this instance is attributed 
to this energy-gap closing. 
That is, the system is in the ground state before the energy-gap closing point.
At this point, however, a nonadiabatic transition to the first excited state occurs.
Consequently, we cannot obtain the ground state at the end.
This time evolution is depicted by the dotted arrows in Fig.~\ref{fig-histogram}(d).

\section{Excited-state AQC with KPOs}
\label{sec-excited-AQC}

\begin{figure*}
	\includegraphics[width=1.8\columnwidth]{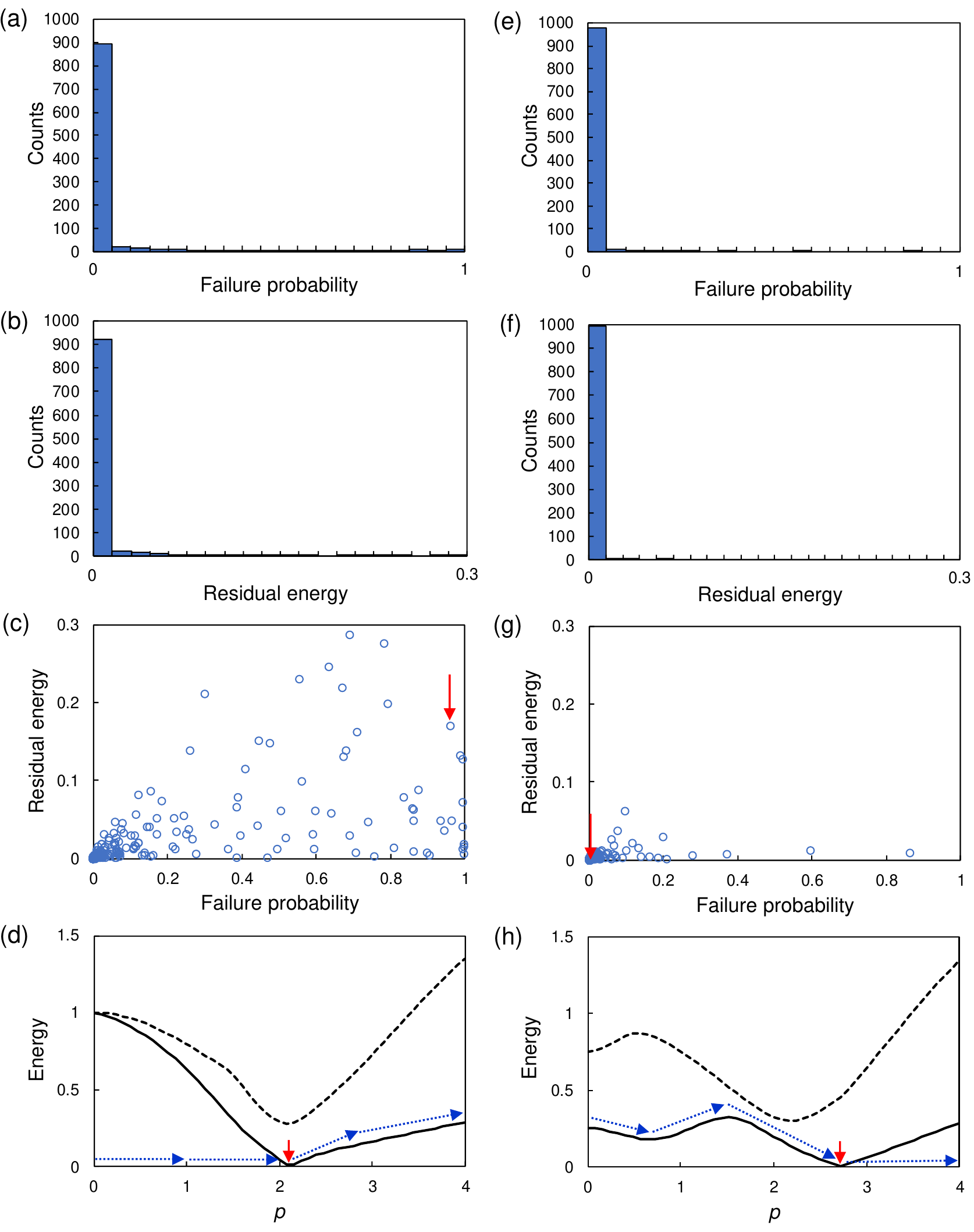}
	\caption{Simulation results for 1000 random instances of the four-spin Ising problem.
	(See Appendix~\ref{sec-simulation} for the details of the simulations.)
	$\{ J_{i,j} \}$ and $\{ h_i \}$ are set randomly from the interval $(-1, 1)$
	and normalized by the maximum magnitude of them.
	(a)--(c) Results for the ground-state AQC with KPOs.
	(d) Energy levels from the ground-state energy as functions of pump amplitude $p$ 
	in the instance indicated by the vertical arrow in (c).
	Solid line: first excited state.
	Dashed line: second excited state.
	(See Appendix~\ref{sec-instance} for the details of the instance.
	Also see Appendix~\ref{sec-energy} for numerical calculation of the energies.)
	Here we use the units that $\hbar=1$ and $K=1$
	($K$ and $\hbar K$ are the units of frequency and energy, respectively).
	Vertical arrow in (d) indicates energy-gap closing. 
	Dotted arrows in (d) depict time evolution of the state during the AQC.
	(e)--(h) Corresponding results for the proposed approach 
	using the excited-state AQC with KPOs.}
	\label{fig-histogram}
\end{figure*}

Here we present our proposed approach.
To set the vacuum state to the first excited state of the initial Hamiltonian,
we set one of the initial detunings, e.g. $\Delta_1^{(0)}$, to a negative value.
Then, the single-photon and two-photon energies for the corresponding KPO at the initial time 
are expressed, respectively, 
as ${\hbar \Delta_1^{(0)}}$ and  ${\hbar (K + 2\Delta_1^{(0)} )}$. 
Thus, the vacuum state is the first excited state when ${-K/2 < \Delta_1^{(0)} <0}$.

In the hard instance discussed in the last section,
its energy levels change from the ones  in Fig.~\ref{fig-histogram}(d) 
to the ones in Fig.~\ref{fig-histogram}(h), 
where one of the initial detunings is set to ${-K/4}$ (the others are $K$). 
From Fig.~\ref{fig-histogram}(h), 
we expect to obtain the ground state via the nonadiabatic transition 
from the first excited state to the ground state
at the energy-gap closing point, 
as depicted by the dotted arrows in Fig.~\ref{fig-histogram}(h).
In fact, our simulation shows that 
the failure probability and the residual energy are improved 
from 0.963 and 0.171 
to ${7.10 \times 10^{-4}}$ and ${6.87 \times 10^{-4}}$, respectively, 
by the excited-state AQC, 
indicated by the vertical arrows in Figs.~\ref{fig-histogram}(d) and \ref{fig-histogram}(h).

Note that the excited-state AQC does not always succeed.
Our proposal is to try the ground-state AQC and the excited-state AQC with 
a negative $\Delta_i^{(0)}$ ($i=1, \ldots, N$)
and to select the best result among the $({N+1})$ cases.
By this approach, 
the results for the 1000 random instances are dramatically improved 
from Figs.~\ref{fig-histogram}(a)--\ref{fig-histogram}(c)
to Figs.~\ref{fig-histogram}(e)--\ref{fig-histogram}(g).
This demonstrates the usefulness of the proposed approach.

\section{Effects of dissipation}
\label{sec-dissipation}

\begin{figure}[t]
	\includegraphics[width=\columnwidth]{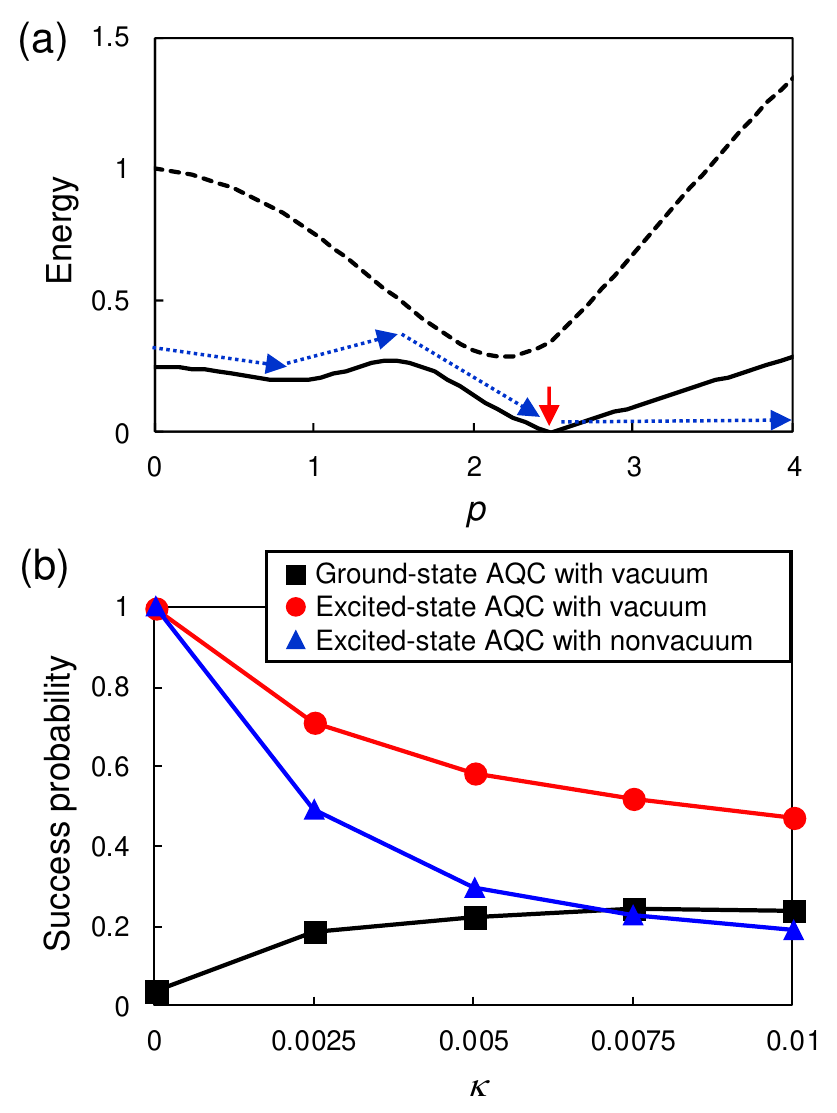}
	\caption{Simulation results for a hard instance in the presence of dissipation.
	The instance is indicated by the vertical arrow in Fig.~\ref{fig-histogram}(c).
	(See Appendix~\ref{sec-instance} for the details of the instance.)
	(a) Energy levels from the ground-state energy as functions of pump amplitude $p$ 
	in the excited-state AQC started with a really excited (nonvacuum) state 
	(see the main text).
	Solid line: first excited state.
	Dashed line: second excited state.
	(See Appendix~\ref{sec-energy} for numerical calculation of the energies.)
	Here, we use the units that $\hbar=1$ and $K=1$
	($K$ and $\hbar K$ are the units of frequency and energy, respectively).
	Vertical arrow in (a) indicates energy-gap closing. 
	Dotted arrows in (a) depict time evolution of the state during the AQC.
	(b) Success probability as a function of decay rate, $\kappa$, for photons in each KPO.
	Squares: ground-state AQC.
	Circles: excited-state AQC started with the vacuum state.
	Triangles: excited-state AQC started with a really excited state.
	See Appendix~\ref{sec-simulation} for the details of the simulations.}
	\label{fig-dissipation}
\end{figure}

The merit of the present approach is 
the robustness of the initial state against errors due to dissipation.
To demonstrate it,
we solved the same instance as above by the ground-state and excited-state AQCs
in the presence of dissipation
(see Appendix~\ref{sec-simulation} for the details of the simulation).
For comparison, we also solved it by the excited-state AQC started with a really excited state,
where one of the initial detunings is set to a positive value 
smaller than the others and the corresponding KPO is initially set in the single-photon state, 
resulting in the situation where the initial state is the first excited state.
The energy levels in the case where one of the initial detunings is set to $K/4$
(the others are $K$) 
are shown in Fig.~\ref{fig-dissipation}(a),
from which we expect to successfully obtain the ground state via
the nonadiabatic transition from the first excited state to the ground state
at the energy-gap closing point, 
as depicted by the dotted arrows in Fig.~\ref{fig-dissipation}(a).

The results for the three cases are shown in Fig.~\ref{fig-dissipation}(b).
(The success probability is one minus the failure probability.)
As expected, 
the excited-state AQC started with a really excited state 
achieved a high success probability (0.9999) in the absence of dissipation
(${\kappa = 0}$).
However, this performance is rapidly degraded as the decay rate $\kappa$ increases.
When $\kappa$ is as large as $0.01K$, 
the excited-state AQC started with a really excited state
becomes worse than the ground-state AQC.
(The enhancement of the performance of the ground-state AQC by dissipation
is explained by quantum heating~\cite{Goto2018a}.)
On the other hand, 
the performance of the excited-state AQC started with vacuum states is more robust,
as shown in Fig.~\ref{fig-dissipation}(b). 
This result demonstrates the robustness of the present approach 
against errors due to dissipation.

\section{Conclusions}
\label{sec-conclusion}

We have proposed a new approach to excited-state AQC,
which is started with not a really excited state, 
but the most stable state, namely, the vacuum state.
This is based on the use of the effective Hamiltonian for a driven quantum system, 
that is, a KPO network,
which allows one to set the vacuum state to an excited state for the Hamiltonian used for AQC.
A hard instance, where the standard ground-state AQC ends up with failure
because of energy-gap closing, can be solved by the excited-state AQC 
exploiting a nonadiabatic transition from the first excited state to the ground state 
at an energy-gap closing point.
Since the excited-state AQC is started with vacuum states,
this AQC is robust against errors due to dissipation,
which has been confirmed by numerical simulations.
Thus, the present approach enhances the power of AQC and, in particular, 
offers a new way for tackling the energy-gap closing problem 
by harnessing a property of driven quantum systems.

\section*{Acknowledgments}

This work was supported by JST ERATO (Grant No. JPMJER1601).

\begin{appendix}

\section{Numerical simulations}
\label{sec-simulation}

To solve the four-spin Ising problem by AQCs with KPOs and 
to obtain the results in Figs.~\ref{fig-histogram} and \ref{fig-dissipation},
we numerically solved the Schr\"odinger equation 
or the master equation with the Hamiltonian in Eq.~(\ref{eq-H}):
\begin{align}
\frac{d}{dt} |\psi \rangle
&=
-\frac{i}{\hbar} H(t) |\psi \rangle,
\label{eq-Schrodinger}
\\
\frac{d}{dt} \rho 
&=
-\frac{i}{\hbar} [H(t), \rho ]
+ \kappa \! \left( 2 a \rho a^{\dagger} - a^{\dagger} a \rho - \rho a^{\dagger} a \right) \!, 
\label{eq-master}
\end{align}
where $|\psi \rangle$ and $\rho$ are the state vector and the density operator, 
respectively, describing the KPO network, 
$[O_1, O_2] = O_1 O_2 - O_2 O_1$ is the commutation relation between $O_1$ and $O_2$,
and $\kappa$ is the decay rate for photons in each KPO.
To solve the master equation numerically, 
we used the quantum-jump approach~\cite{Scully,Breuer,Plenio1998a},
where instead of the master equation (\ref{eq-master}), 
we solve the Schr\"odinger equation with the non-Hermitian Hamiltonian:
\begin{align}
\frac{d}{dt} |\psi \rangle
&=
-\frac{i}{\hbar} H'(t) |\psi \rangle,
\quad
H'(t) = H(t) - i \hbar \kappa a^{\dagger} a.
\label{eq-jump}
\end{align}
The results in Fig.~\ref{fig-dissipation}(b) 
were obtained by taking the averages over 1000 trials 
of the Monte-Carlo simulation in the quantum-jump approach.

In the numerical simulations,
we truncated the Hilbert space for each KPO at a maximum photon number of 14, 
and represented the state vector $|\psi \rangle$ in the photon-number basis.
The resultant differential equations were solved 
by the fourth-order Runge-Kutta method with a time step of ${1/(500K)}$.

In all the simulations, 
the parameters were set as follows (if not mentioned):
\begin{align}
p(t)&=
p_f \sin \! \frac{\pi t}{2T}, \quad p_f=4K,
\\
\Delta_i \! (t)&=
\Delta_i^{(0)} \! \cos \! \frac{\pi t}{2T}, \quad \Delta_i^{(0)}=K,
\\
\xi(t)&=
\xi_f \sin \! \frac{\pi t}{2T}, \quad \xi_f = \frac{K}{4},
\end{align}
where ${T=400/K}$ is the computation time of the AQCs.

\section{Details of the hard instance}
\label{sec-instance}

The coupling coefficients and the local fields
of the instance used for Figs.~\ref{fig-histogram}(d), \ref{fig-histogram}(h), and \ref{fig-dissipation}
[indicated by the vertical arrow in Fig.~\ref{fig-histogram}(c)] 
are defined as follows:
\begin{align}
& J_{1,2}=J_{2,1}=0.266654, \nonumber \\
& J_{1,3}=J_{3,1}=0.886155, \nonumber \\
& J_{1,4}=J_{4,1}=0.019833, \nonumber \\
& J_{2,3}=J_{3,2}=0.071362, \nonumber \\
& J_{2,4}=J_{4,2}=-0.446931, \nonumber \\
& J_{3,4}=J_{4,3}=-1, \nonumber \\
& h_1=-0.340697, \nonumber \\
& h_2=-0.546404, \nonumber \\
& h_3=0.501731, \nonumber \\
& h_4=-0.296651. \nonumber
\end{align}

\begin{figure}[t]
	\includegraphics[width=\columnwidth]{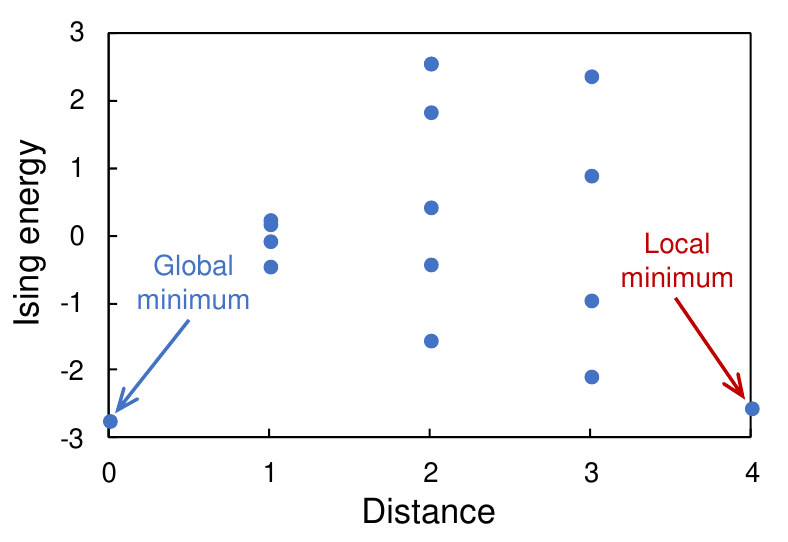}
	\caption{Energy landscape of the hard instance 
	[indicated by the vertical arrow in Fig.~\ref{fig-histogram}(c)].
	``Distance" is the Hamming distance 
	between the optimal solution (the global minimum) 
	and each configuration~\cite{comment-Hamming}.}
	\label{fig-landscape}
\end{figure}

The energy landscape of this instance is depicted in Fig.~\ref{fig-landscape}.
It is found that there is a nonglobal local minimum far from the global minimum.
This may be the reason why this instance is hard.

\section{Numerical calculation of energies}
\label{sec-energy}

The energies in Figs.~\ref{fig-histogram}(d), \ref{fig-histogram}(h), and \ref{fig-dissipation}(a) 
are obtained by diagonalizing the Hamiltonian matrix.
However, the size of this matrix (${15^4 \times 15^4}$ in the present case) 
is too large to directly diagonalize.
Instead of the direct diagonalization,
we calculated the energies as follows.

We first numerically obtain the eigenvectors for each KPO
by diagonalizing each term in the first term of $H(t)$ in Eq.~(\ref{eq-H}).
Taking $N_e$ eigenvectors from low energies as a basis,
we obtain a ${N_e^4 \times N_e^4}$ matrix representation of $H(t)$.
Finally, we diagonalize this matrix and obtain the energies.
Note that we can obtain the exact diagonalization if we take all the basis vectors ($N_e=15$).
To reduce the computational costs, 
we set ${N_e=6}$ in the present calculations, 
because the energies obtained sufficiently converge.
This approach based on the low-energy approximation 
is valid when $\xi$ is small compared to $K$, 
as in the present case.

\end{appendix}


\begin{thebibliography}{99}
\bibitem{comment-xi-A}
$\xi$ has the dimension of frequency and $A$ is dimensionless.
\bibitem{comment-coherent}
A coherent state ${|\alpha \rangle }$ is defined as 
an eigenstate of the annihilation operator $a$: 
$a {| \alpha \rangle} = \alpha {| \alpha \rangle}$.
\bibitem{comment-Hamming}
The Hamming distance between two spin configurations, $\{ s_i \}$ and $\{ s'_i \}$,
is defined by the number of spin pairs, $(s_i, s'_i)$, satisfying $s_i \neq s'_i$.
\end{thebibliography}
\end{document}